\documentclass[onecolumn,showpacs,preprintnumbers,amsmath,amssymb]{revtex4}
\usepackage{graphicx}
\usepackage{graphics}
\usepackage{amssymb}
\usepackage{pifont}
\usepackage{txfonts}
\usepackage{bm}

\newcommand{\ie}{{\it i.e.}}
\newcommand{\eg}{{\it e.g.}}
\begin{document}

\title{Manipulation of atom-to-molecule conversion in a magnetic lattice}

\author{Ning-Ju Hui${}^1$, Li-Hua Lu${}^{1*}$, Li-Bin Fu${}^2$ and You-Quan Li${}^1$}
\affiliation{${}^1$Department of Physics, Zhejiang University, Hangzhou 310027, China\\
${}^2$ Institute of Applied Physics and Computational
Mathematics, Beijing 100088, China}

\begin{abstract}{\bf Abstract:} The atom-to-molecule conversion by the technique of optical Feshbach resonance in a magnetic lattice
is studied in the mean-field approximation. For the case of shallow lattice, we give the dependence of the atom-to-molecule conversion efficiency on the tunnelling strength and the atomic interaction by taking a double-well as an example. We find that one can obtain a high atom-to-molecule conversion by tuning the tunnelling and interaction strengths of the system. For the case of deep lattice, we show that the existence of lattice can improve the atom-to-molecule conversion for certain initial states.

\end{abstract}

\received{\today}
\maketitle

\section{Introduction}

Since Bose-Einstein condensations (BECs) in dilute atomic gases were realized in 1995,
the study of cold atoms becomes  a remarkable research area which has been extended from atomic  to
molecular systems in recent years.
Molecular BECs are versatile not only for cold atomic physics but
also for other research areas~\cite{novel physical phenomena 1_BEC-BCS crossover,novel physical phenomena 2_ultracold chemistry,high-precision measurements}, because they  include more degrees of freedom than atomic systems. To realize
molecular BECs, one usually  converts ultracold atoms into molecules through resonant photoassociation (optical Feshbach resonance) or magnetoassociation (magnetic Feshbach resonance)~\cite{MBEC creation1 by FR,MBEC creation2 by FR,MBEC creation3 by FR,MBEC creation4 by FR,MBEC creation5 by FR,FR3,FR4,FR5} rather than cool molecules directly. Additionally, besides the technique of Feshbach resonance, the stimulated Raman adiabatic passage technique  was also used in the conversion of Bose-Fermi mixture into molecules~\cite{stimulated Raman adiabatic passage technique2,stimulated Raman adiabatic passage technique3}.
  In the above works, the  atom-to-molecule conversion system were all confined in a single well. Note that in recent experiments~\cite{increase the molecule lifetime1,increase the molecule lifetime2,atom-molecule-OL1,atom-molecule-OL2}, the  atom-to-molecule conversion system  confined in an optical lattice was also studied, where the atom-to-molecule conversion efficiency and the lifetime of molecules can be improved   due to the suppression of the inelastic collisions.  Similar to optical lattice, magnetic lattice can also be expected to improve the atom-to-molecule conversion since ultracold atoms have been  successfully  transferred to the magnetic lattice potential experimentally in 2008~\cite{exp-magnetic-lattice}. Compared with optical lattice, magnetic lattice has several distinct advantages, such as high stability with low technical noise and low heating rates, large and controllable barrier heights and so on. It is thus meaningful to study the property of atom-to-molecule conversion in a magnetic lattice.

 In this paper, we consider an atom-to-molecule conversion system in a magnetic lattice. In the mean field approximation, we study the influence of magnetic lattice on the atom-to-molecule conversion efficiency
 and show how to improve the atom-to-molecule conversion in a magnetic lattice. The rest of this paper is organized as follows. In Sect.~\ref{sec:model}, we present the model and the dynamical equations. In Sect.~\ref{sec:J unequal 0}, we consider the case of shallow lattice, \ie, the atoms can  tunnel between the nearest neighbouring sites. Taking a double-well as an example, we study the time evolution of molecular density and the effect of parameters of the system on the atom-to-molecule conversion. We also  confirm our numerical results with the help of fixed points and energy contours of the system. In Sect.~\ref{sec:J=0}, we consider deep lattice, \ie, the atoms can not tunnel between the lattice sites. We show that the existence of the lattice can improve the atom-to-molecule conversion for certain initial states. In the last section, we give a brief summary and discussion.

\section{Model and general formulation}\label{sec:model}
We consider the atom-to-molecule conversion via optical Feshbach resonance in the magnetic lattice, where the atoms are subject to the
lattice but the molecules are not. The Hamiltonian describing such a system can be written as,
\begin{eqnarray}
\label{eq:Hamiltonian}
 \hat{H_l}&=&-J\sum_{\langle i,j\rangle}(\hat{a}^{\dag}_i\hat{a}_j+\hat{a}^{\dag}_j\hat{a}_i)
 +\frac{\widetilde{U}_a}{2}\sum_{i}\hat{n}_{ai}(\hat{n}_{ai}-1)
 +\frac{\widetilde{U}_b}{2}\hat{n}_{b}(\hat{n}_{b}-1)
 +\frac{\tilde{g}}{2}\sum_{i}(\hat{b}^{\dag}\hat{a}_i\hat{a}_i+\hat{b}\hat{a}^{\dag}_i\hat{a}^{\dag}_i)
 +\delta\hat{b}^{\dag}\hat{b},
\end{eqnarray}
where the operator $\hat{a}_{i(j)}^{\dag}$ and $\hat{a}_{i(j)}$ creates and annihilates a bosonic atom in the $i(j)$th  site separately and $\langle i,j\rangle$ denotes the two nearest neighbouring lattice sites, while $\hat{b}^{\dag}$ and $\hat{b}$ creates and annihilates a bosonic molecule. They obey the commutation relation $[\hat{a}_{i}, \hat{a}_{j}^{\dag}]=\delta_{ij}$ and $[\hat{b},\hat{b}^{\dag}]=1$. Here the parameter $J$ denotes the tunneling strength of atoms
between the nearest neighbouring sites, which can be tuned by changing the
distance between the neighbouring lattice sites or the height of the
potential barrier separating the neighbouring lattice sites. $\widetilde{U}_a$ and $\widetilde{U}_b$ is the interaction strength between the atoms and molecules, respectively. As we know, $U_a$ can be written as $U_a=2\pi\hbar^2a/\mu$, where $a$ and $\mu$ are scattering length and reduced mass, respectively. The scattering length can be tuned by the optical Feshbach resonance technique. $\delta$ is the energy detuning between the atomic and molecular states, and $\tilde{g}$ refers to the coupling strength between atoms and molecules.
The molecules in spin singlet state ($S=0$) are not subject to the magnetic lattice, so there is no subscript for $b$ in the Hamiltonian (\ref{eq:Hamiltonian}).

In order to study the property of atom-to-molecule conversion, we need to give the dynamical equations of the
system. The classical field description is an excellent approximation if the quantum fluctuation is small. As we know, the magnitude of quantum fluctuation around the condensate state scales down as $1/\sqrt{N}$ in zero temperature with $N$ the particle number. There are usually $10^4-10^7$ particles in dilute BEC experiments, so we adopt the mean-field description~\cite{mean-field}.
In the mean field approximation, by replacing operators with their expectation values, \ie, $\hat{a}_{i(j)}\rightarrow\langle\hat{a}_{i(j)}\rangle=\tilde{a}_{i(j)}$ and $\hat{b}\rightarrow\langle\hat{b}\rangle=\tilde{b}$, one can easily give the dynamical equations for $\tilde{a}_{i(j)}$  and $\tilde{b}$ with the help of Heisenberg equations of motion for
$\hat{a}_{i(j)}$ and $\hat{b}$. These equations obey the conservation law $\sum^{}_i\hat{a}^\dag_i\hat{a}_i+2\hat{b}^{\dag}\hat{b}=N$. To simplify the
calculation, one usually assumes
 $a_{i(j)}=\tilde{a}_{i(j)}/\sqrt{N}$ and $b=\tilde{b}/\sqrt{N}$ with $\sum^{}_i|a_i|^2+2|b|^2=1$. Then the dynamical equations for $a_{i(j)}$ and $b$ can be rewritten as,
\begin{eqnarray}
\label{eq:evolution equation of optical lattice}
 i\dot{a}_i&=&-J(a_{i+1}+a_{i-1})+U_a|a_i|^2a_i+gba^*_i,\nonumber\\
 i\dot{b}&=&U_b|b|^2b+\delta b+\frac{g}{2}\sum^{}_ia^2_i,
\end{eqnarray}
where $U_a=\widetilde{U}_aN$, $U_b=\widetilde{U}_bN$, $g=\tilde{g}\sqrt{N}$ and natural units has been used. Note that the interaction strength between molecules is much
smaller than that between atoms in most experiments, so we ignore the interaction
between molecules in the following calculation, \ie, $U_{b}=0$. Meanwhile the energy detuning is chosen as $\delta=0$ .
We study the atom-to-molecule conversion efficiency, which is defined as twice the density of the largest molecules ($2|b|^2_{max}$) in the time evolution, for the cases of shallow lattice and deep lattice in the following two sections separately.

\section{Shallow lattice}\label{sec:J unequal 0}
In this section, we consider that the atoms tunnel between the nearest neighbouring sites, \ie, $J\neq 0$. Although we can give the dynamical
properties of the system for different number of lattice size $L$ by solving
(\ref{eq:evolution equation of optical lattice}) numerically,
here we mainly take double-well as a typical example. For the case
of double-well, Eqs.~(\ref{eq:evolution equation of optical
lattice}) is simplified to,
\begin{eqnarray}
\label{eq:evolution equation of double well}
 i\dot{a}_1&=&-Ja_{2}+U_a|a_1|^2a_1+gba^*_1,\nonumber\\
 i\dot{a}_2&=&-Ja_{1}+U_a|a_2|^2a_2+gba^*_2,\\
 i\dot{b}&=&g(a^2_1+a^2_2)/2,\nonumber
\end{eqnarray}
where $\delta=0$ and $U_b=0$ have been taken.
Tunnelling strength $J$, atomic interaction strength $U_a$ and the atom-molecule coupling strength $g$ have the same dimension.
We choose $g$ as unity and then all quantities are renormalized to be dimensionless.

\subsection{Evolution and Conversion Efficiency}

\begin{figure}[tbph]
\centering
\includegraphics[width=140mm]{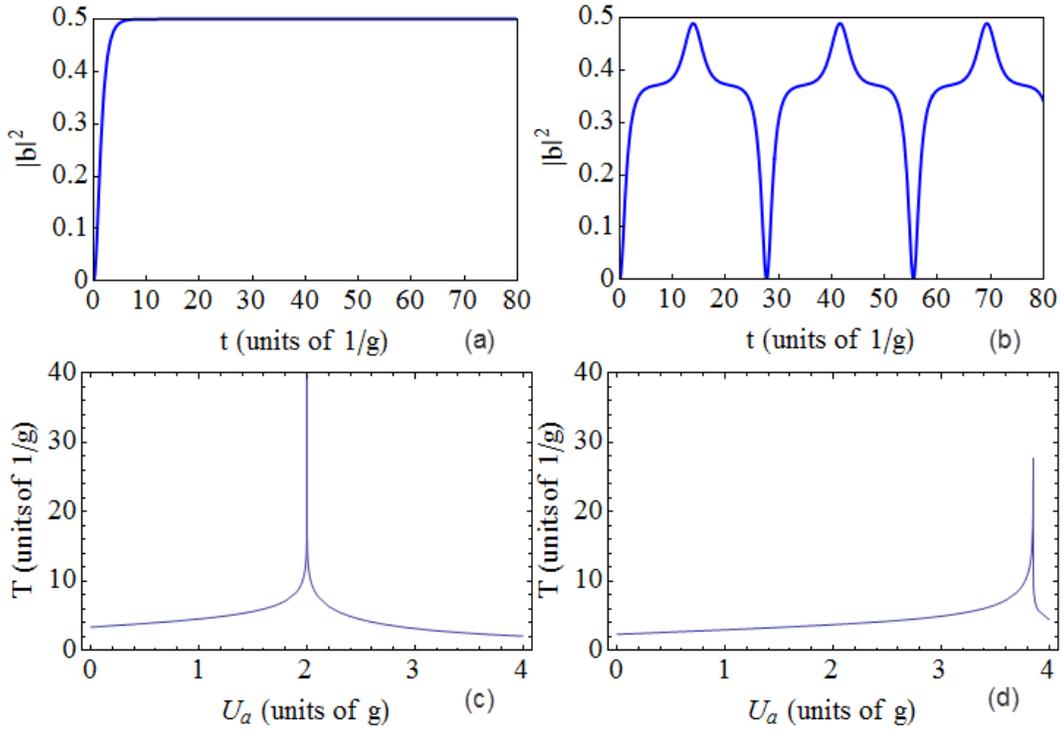}
\caption{(Colour online) Time dependence of the molecular density with symmetrical
initial conditions in (a) and (b).
The parameters are $J=0.5g$ and $U_a=2g$ (a), $J=g$, $U_a=3.853g$
(b). The period of Rabi oscillation $T$ versus $U_a$ in (c) and (d). The parameters are
$J=0.5g$ (c) and $J=g$ (d).}
\label{fig:completed_and_uncompleted_evolution}
\end{figure}
Now we are in the position to study the atom-to-molecule conversion of the BECs in double wells with the help of (\ref{eq:evolution
equation of double well}). Here we consider the symmetrical
initial state $a_{1,2}=\sqrt{1/2}$ and $b=0$. We plot the time
dependence of molecular density for different parameters in
Fig.~\ref{fig:completed_and_uncompleted_evolution} (a), (b) and
the period of Rabi oscillation~\cite{period of Rabi oscillation} versus $U_a$ in
Fig.~\ref{fig:completed_and_uncompleted_evolution} (c), (d). Fig.~\ref{fig:completed_and_uncompleted_evolution} (a) shows that all atoms can be converted into
molecules and the system will always stay in the pure
molecular state, \ie, the oscillation period for this condition is
infinite which corresponds to the peak in Fig.~\ref{fig:completed_and_uncompleted_evolution} (c). Whereas, from
Fig.~\ref{fig:completed_and_uncompleted_evolution} (b), not all of atoms can be converted into molecules for some parameters
and the system oscillates periodically are found. The oscillation period for this condition is
finite corresponding with the peak in Fig.~\ref{fig:completed_and_uncompleted_evolution} (d). The above results enlighten us on choosing the appropriate parameters for high atom-to-molecule conversion efficiency. So it is necessary to find the suitable relation between tunnelling strength $J$ and atomic interaction strength $U_a$.
\begin{figure}[tbph]
\centering
\includegraphics[width=140mm]{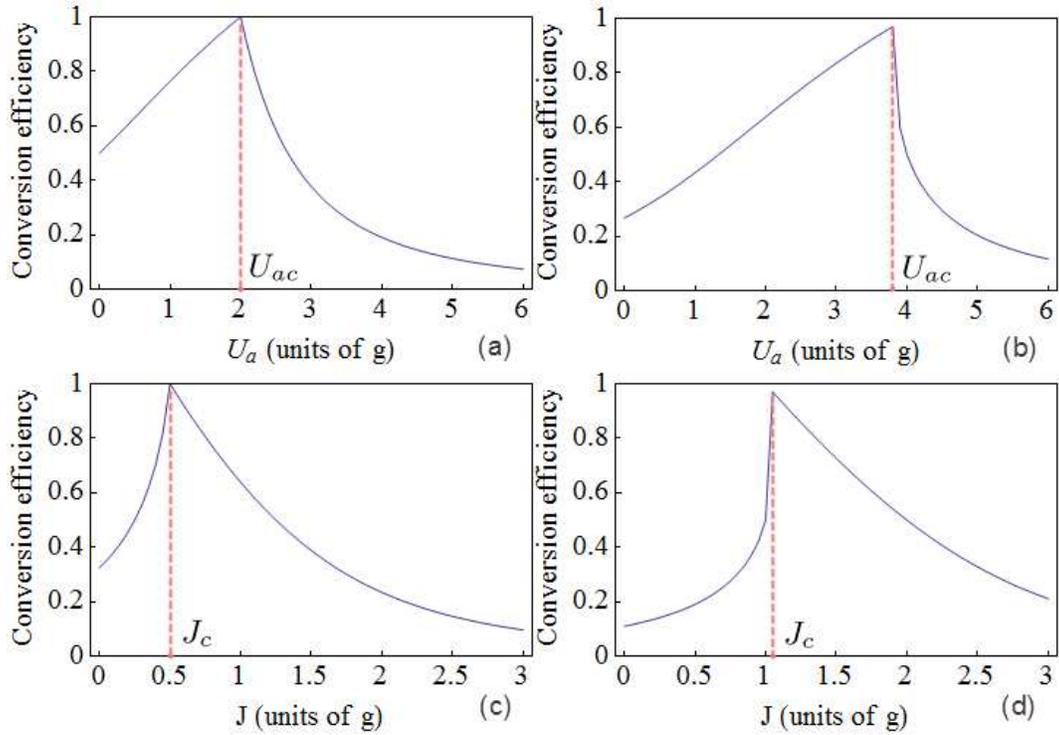}
\caption{(Colour online) Dependence of the atom-to-molecule conversion efficiency on the atomic interaction strength (a) , (b) and the tunnelling strength (c), (d) for double-well. The parameters are $J=0.5g$ (a),  $J=g$ (b),  $U_a=2g (c)$, and  $U_a=4g (d)$. The vertical pink dashed lines are guides to the eye.
} \label{fig:completed_and_uncompleted_conversion_efficiency}
\end{figure}

 We plot the dependence of atom-to-molecule conversion efficiency on the atomic interaction strength in  Fig.~\ref{fig:completed_and_uncompleted_conversion_efficiency} (a) and (b),
 and on the tunnelling strength in Fig.~\ref{fig:completed_and_uncompleted_conversion_efficiency} (c) and (d).
 Fig.~\ref{fig:completed_and_uncompleted_conversion_efficiency} (a) and (b) show that for $J\neq0$,
 the increase of interaction strength can improve the atom-to-molecule conversion when $U_a$ is smaller than the critical value $U_{ac}$,
 while suppress the atom-to-molecule conversion when $U_a>U_{ac}$,
 which is different from the case of $J=0$ plotted in Fig.~\ref{fig:comparison-with-single-well} (b).
 Additionally, for a given value of $J$, when $U_a=U_{ac}$,
 one can get the highest atom-to-molecule conversion efficiency which is dependent on the value of $J$.
 Comparing Fig.~\ref{fig:completed_and_uncompleted_evolution} (c) and
 (d) with Fig.~\ref{fig:completed_and_uncompleted_conversion_efficiency} (a) and
 (b), we find that both
 the periods of Rabi oscillation versus $U_a$ and the dependence of the maximum
 atom-to-molecule conversion efficiency on $U_a$ exhibit a similar behaviour and they have the same parameters of the peaks, \eg, $J=0.5g$, $U_a=2g$ for Fig.~\ref{fig:completed_and_uncompleted_evolution} (c) and Fig.~\ref{fig:completed_and_uncompleted_conversion_efficiency} (a) and
 $J=g$,  $U_a=3.853g$ for Fig.~\ref{fig:completed_and_uncompleted_evolution} (d) and Fig.~\ref{fig:completed_and_uncompleted_conversion_efficiency} (b).
When $J=0.5g$ and $U_a=2g$, all of atoms can be converted into
molecules and stay in the pure molecular state finally with infinite
period. While only partial atoms can be converted into
molecules and the period is finite with parameters $J=g$ and
$U_a=3.853g$.
Fig.~\ref{fig:completed_and_uncompleted_conversion_efficiency} (c) and (d) tell us that for the case of $U_a\neq 0$, the dependence of atom-to-molecule conversion efficiency on the tunnelling strength  does not change monotonically but  there is a critical value $J_c$ where the atom-to-molecule conversion efficiency is the highest.

\begin{figure}[tbph]
\centering
\includegraphics[width=140mm]{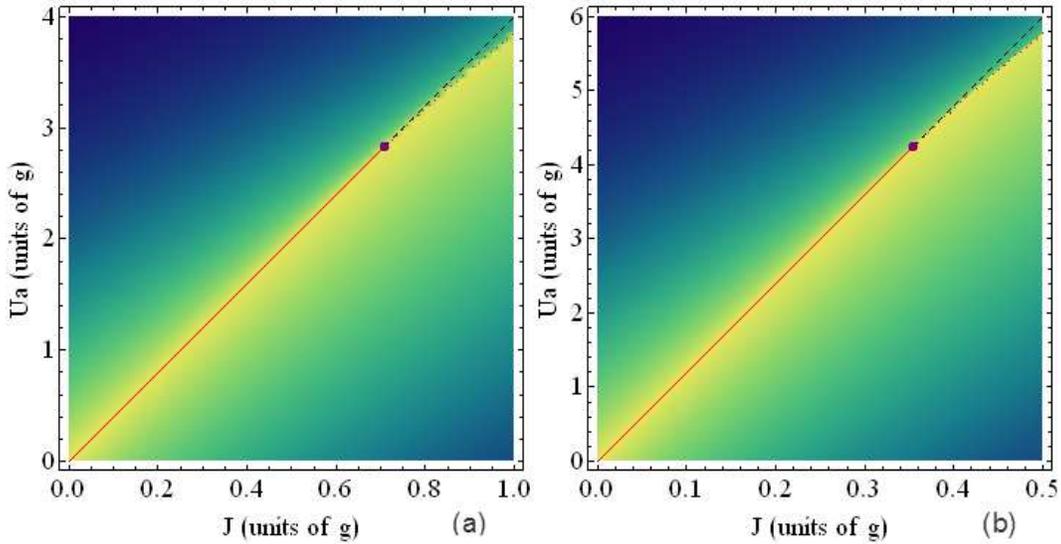}
\caption{(Colour online) Projection of the atom-to-molecule conversion efficiency in the $J$-$U_a$ plane for the cases of double-well (a) and triple-well (b). The red solid line and red dotted line distinguish the complete and uncomplete conversion respectively. The purple filled points are the general critical point. The black dashed lines are $U_a=4J$ (a) and $U_a=12J$ (b) which are guides to the eye.} \label{fig:critical_behavior}
\end{figure}

In order to give further overall dependence of atom-to-molecule conversion on the strengths of tunnelling and interaction of the system, we plot
the projection of the atom-to-molecule conversion efficiency in the $J$-$U_a$ plane in Fig.~\ref{fig:critical_behavior} (a). The purple filled point $(\mathcal{J}_c, \mathcal{U}_{ac})=(g/\sqrt{2},2\sqrt{2}g)$ in Fig.~\ref{fig:critical_behavior} (a) is
an important critical point called ``general critical point". When $J<\mathcal{J}_c$, all of atoms can be converted into molecules if the parameters are selected as the red solid line, \ie, $U_a=4J$. On the other hand, for the case of  $J>\mathcal{J}_c$, although not all of atoms can be converted into molecules no matter what parameters are taken, one can also  get higher atom-to-molecule conversion by taking the parameters on the red dotted line which deviates from the line $U_a=4J$ (\ie, the black dashed line). Fig.~\ref{fig:critical_behavior} shows both the atomic interaction and the tunnelling strength can affect the atom-to-molecule conversion efficiency and one can get high conversion by tuning the parameters of the systems properly. In order to convert all atoms into molecules, one should choose the parameters on the red solid line, \ie,  $U_a=4J$ and $J<\mathcal{J}_c$.

Additionally, we extend our study to triple-well case and give the projection of the atom-to-molecule conversion efficiency in the $J$-$U_a$ plane in Fig.~\ref{fig:critical_behavior} (b). There is also a general critical point $(\mathcal{J}_c, \mathcal{U}_{ac})=(g\sqrt{2}/4,3\sqrt{2}g)$ and the relation of the red solid line is $U_a=12J$ as $J<\mathcal{J}_c$. The discussion of the dependence of atom-to-molecule conversion efficiency on $U_a$, $J$ and the oscillated period are similar to the double-well case.

\subsection{Fixed Points and Energy Analysis}

In order to understand the dependences of the
conversion efficiency on the parameters $J$ and $U_a$, we study the
property of the fixed points.  By expressing
$a_{i}=\sqrt{\rho_{ai}}e^{\theta_{ai}}$ and $b
=\sqrt{\rho_b}e^{\theta_b}$, we can write Eq.~(\ref{eq:evolution
equation of double well}) as
\begin{eqnarray}
\label{eq:Hamiltonian double-well}
\dot{z}&=&-2J\sqrt{\rho_{a1}\rho_{a2}}\sin(2\phi_a)-g\sqrt{\rho_b}[\rho_{a1}\sin(\phi-2\phi_a)
-\rho_{a2}\sin(\phi+2\phi_a)],\nonumber\\
\dot{\phi_a}&=&\frac{Jz}{\sqrt{\rho_{a1}\rho_{a2}}}\cos(2\phi_a)+\frac{g}{2}\sqrt{\rho_b}[\cos(\phi-2\phi_a)-\cos(\phi+2\phi_a)]
+\frac{U_a}{2}z,\nonumber\\
\dot{\rho_b}&=&\frac{g}{2}\sqrt{\rho_b}[\rho_{a1}\sin(\phi-2\phi_a)+\rho_{a2}\sin(\phi+2\phi_a)],\\
\dot{\phi}&=&\frac{2J(1-2\rho_b)}{\sqrt{\rho_{a1}\rho_{a2}}}\cos(2\phi_a)-g\sqrt{\rho_b}[\cos(\phi-2\phi_a)+\cos(\phi+2\phi_a)]
-U_a(1-2\rho_b)\nonumber\\
&&+\frac{g}{4\sqrt{\rho_b}}[\rho_{a1}\cos(\phi-2\phi_a)+\rho_{a2}\cos(\phi+2\phi_a)],\nonumber\\
\end{eqnarray}
where $z=\rho_{a1}-\rho_{a2}$, $\phi_a=(\theta_{a2}-\theta_{a1})/2$,  $\phi=\theta_{a1}+\theta_{a2}-\theta_b$ and $\rho_{a1}=(1-2\rho_b+z)/2$, $\rho_{a2}=(1-2\rho_b-z)/2$.
$z$ and $\phi_a$, $\rho_b$ and $\phi$ are mutually canonical conjugations, respectively.
They satisfy
the Hamiltonian canonical equation, $\ie$, $\dot{z}=-\partial H_{cl}/\partial\phi_a$, $\dot{\phi_a}=\partial H_{cl}/\partial z$, $\dot{\rho_b}=-\partial H_{cl}/\partial\phi$ and $\dot{\phi}=\partial H_{cl}/\partial \rho_b$.
The corresponding classical Hamiltonian of the system is  given by
\begin{eqnarray}
\centering
\label{eq:Hamiltonian classical}
 H_{cl}&=&-J\sqrt{(1-2\rho_b)^2-z^2}\cos(2\phi_a)+\frac{U_a}{4}[(1-2\rho_b)^2+z^2]
+\frac{g}{2}\sqrt{\rho_b}[\rho_{a1}\cos(\phi-2\phi_a)+\rho_{a2}\cos(\phi+2\phi_a)].
\end{eqnarray}
We should note that the time evolution of system can be determined by the classical Hamiltonian due to the energy conservation law.

In order to get the fixed point solutions, we set $\dot{z}=0$, $\dot{\phi_a}=0$, $\dot{\rho_b}=0$ and $\dot{\phi}=0$ in Eq.~(\ref{eq:Hamiltonian double-well}). For the initial states $z=0$, $\phi_a=0$,  the atomic distribution and phase in the double wells are the same at any time, \ie, $z(t)=0$ and $\phi_a(t)=0$. Then the conditions for the  fixed points can be written as,
\begin{eqnarray}
\label{eq:fixed points 1}
 g\sqrt{\rho_b}(1-2\rho_b)\sin\phi=0,\nonumber\\
 2J-U_a(1-2\rho_b)+\frac{g(1-6\rho_b)}{2\sqrt{\rho_b}}\cos\phi=0.
\end{eqnarray}
Here we do not write the fixed point solutions in order to save space, although we can get such  solutions through solving Eq.~(\ref{eq:fixed points 1}). Note that when $0\leq J\leq g/\sqrt{2}$, there is a fixed point $\rho_b=1/2$ with $\phi$ being not well defined. Since the fixed point $\rho_b=1/2$ exists for the case of $0\leq J\leq g/\sqrt{2}$, it is possible for the system to reach it, \ie, all of atoms can be converted into molecules by tuning the parameters of the system properly. According to Eq.~(\ref{eq:Hamiltonian classical}), we know that the energy of the system is $-J+U_a/4$ for the initial state we considered, and zero for the fixed point $\rho_b=1/2$.
 So if the system can reach the fixed point  $\rho_b=1/2$, the parameters must satisfy $-J+U_a/4=0$ due to the existence of  energy conservation law. It explains well why the conversation efficiency  corresponding to the parameters taken on the red line in Fig.~\ref{fig:critical_behavior} is 1. Whereas, for the case of $J>g/\sqrt{2}$, the fixed point $\rho_b=1/2$ disappears, and then not all of atoms can be converted into molecules no matter what atomic interaction strength is taken, which is confirmed by Fig.~\ref{fig:completed_and_uncompleted_conversion_efficiency} (b).

\begin{figure}[tbph]
\centering
\includegraphics[width=140mm]{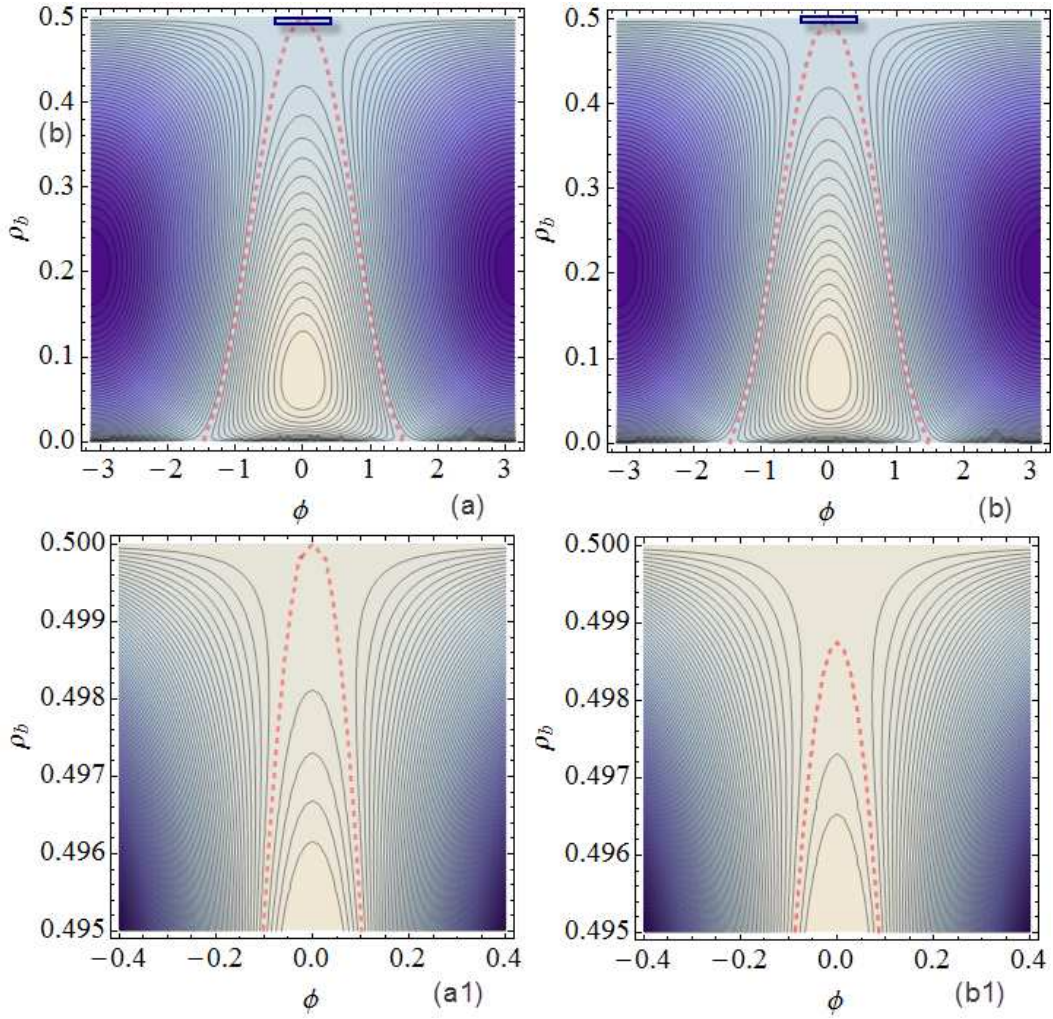}
\caption{(Colour online)  Energy contour in the phase space of $\phi$
and $\rho_b$. The parameters are  $J=0.707 $ and $U_a=2.828$ (a),
$J=0.708$ and $U_a=2.832$ (b). The enlargements of the small blue boxes
in (a) and (b) are (a1) and (b1), respectively. The pink dashed curves
correspond to the energy with the initial states $a_1=a_2=1/\sqrt2$
and $b=0$.} \label{fig:energy_contour}
\end{figure}

In Fig.~\ref{fig:energy_contour},  we  plot the energy contours in the phase space of $\rho_b$ and $\phi$  for the two cases of $J<g/\sqrt{2}$ and $J>g/\sqrt{2}$, respectively. From Fig.~\ref{fig:completed_and_uncompleted_evolution} (a) and (b), we can see that, for a fixed value of $J$, one can get highest atom-to-molecule conversion efficiency when $U_a=U_{ac}$. So in Fig.~\ref{fig:energy_contour},  we take the atomic interaction strength as $U_{ac}$ which is determined by $J$.  Note that  we can obtain $U_{ac}=4J$ analytically for $J<g/\sqrt{2}$,  while the value of $U_{ac}$ needs to be determined by numerical method for $J>g/\sqrt{2}$. For the initial states we considered, the system evolves along the pink dashed curve. Fig.~\ref{fig:energy_contour} (a1) and (b1) further confirm that the system can reach the state $\rho_b=1/2$ for $J<g/\sqrt{2}$ but can not for $J>g/\sqrt{2}$ clearly.

We also study the
property of the fixed points to confirm the dependence of the
conversion efficiency on the parameters $J$ and $U_a$ for the triple-well from Fig.~\ref{fig:critical_behavior} (b) and they are in good agreement. The fixed point $\rho_b=1/2$ exists for the case of $0\leq J\leq g\sqrt{2}/4$ when the atoms can be converted into molecules completely with corresponding critical interaction strength, while the fixed point $\rho_b=1/2$ disappears for the case of $J>g\sqrt{2}/4$ when only part of the atoms can be combined to molecules.

\section{Deep lattice}\label{sec:J=0}

\begin{figure}[tbph]
\centering
\includegraphics[width=140mm]{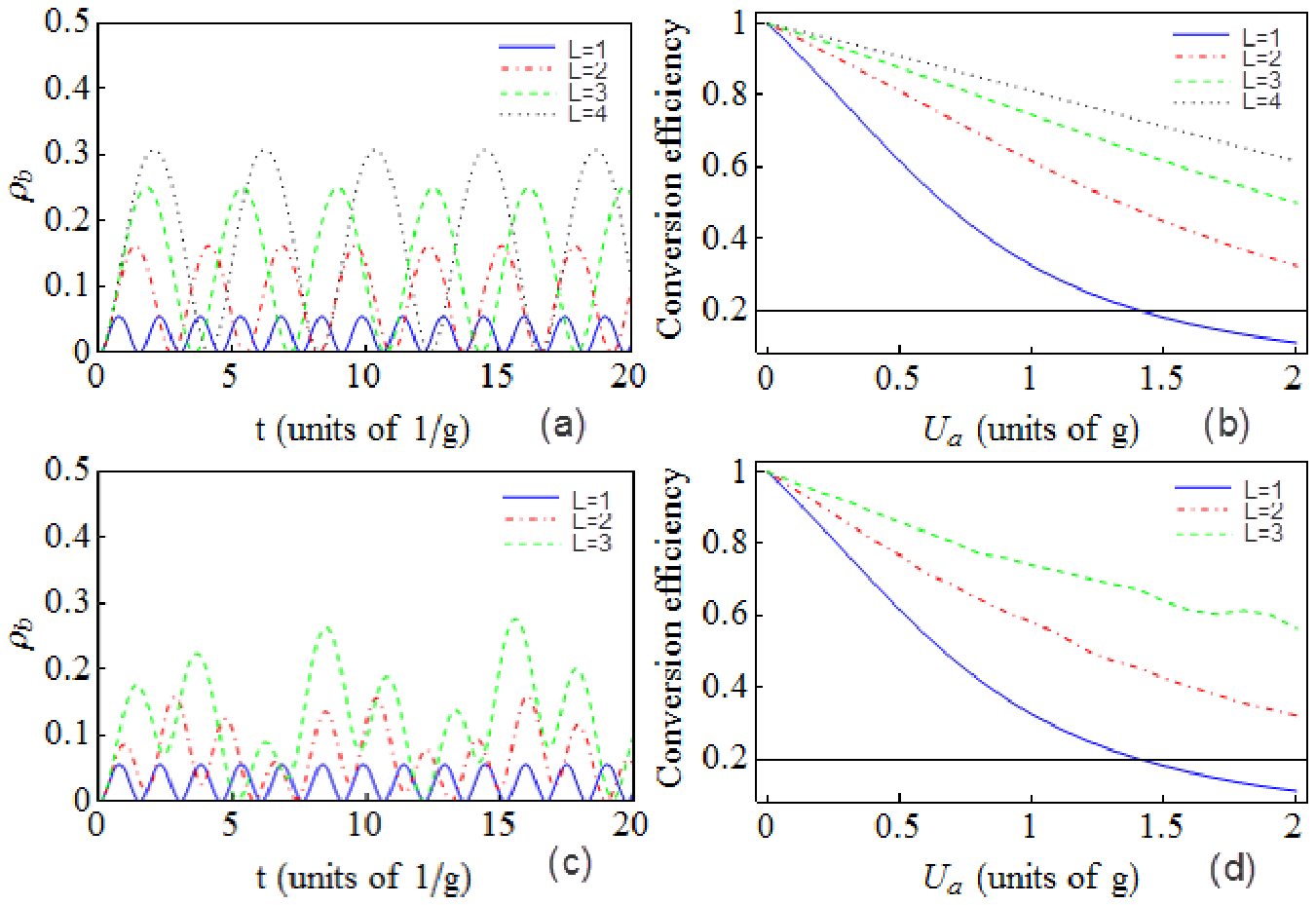}
\caption{(Colour online)  Time dependence of molecular density for
different values of $L$ with symmetrical initial state (a) and asymmetrical initial state (c), dependence of the
atom-to-molecule conversion efficiency on the atomic interaction
strength $U_a$ for different $L$ with symmetrical initial state (b) and asymmetrical initial state (d). The  parameters are $U_a=2g$ and  $J=0$
(a) and (c), $J=0$ (b) and (d).
The asymmetrical initial conditions considered are $|a_1(0)|^2=1/4$, $|a_2(0)|^2=3/4$, $|b(0)|^2=0$ for $L=2$ and $|a_1(0)|^2=1/4$, $|a_2(0)|^2=1/4$, $|a_3(0)|^2=1/2$, $|b(0)|^2=0$ for $L=3$.
 } \label{fig:comparison-with-single-well}
\end{figure}
In this section, we
consider the case of deep lattice, where the height of potential
barrier separating the neighbouring sites is so high that we can ignore the
tunnelling  of atoms between neighbouring site, \ie, $J=0$. For a deep lattice without atom-to-molecule conversion, the ground state has a fixed number of particles per lattice site, and the relative phase between lattice sites is smeared out. However, the existence of atom-to-molecule conversion, where the molecule are not subject to the lattice, makes the phases between lattice sites related and well defined. In order
to study the atom-to-molecule conversion efficiency, we assume there
is no molecule in the system at the initial time and the atoms are
equally populated  in the lattice, \ie, the initial state
is $b=0$ and $a_i=\sqrt{N/L}$ with $L$ being the number of
lattice. Once the initial state is fixed, we can get the time
evolution of the system through solving Eq.~(\ref{eq:evolution
equation of optical lattice}) numerically and the dependence of the atom-to-molecule conversion efficiency on the atomic interaction strength $U_a$ if the value of $L$ is not
very large.  We summarize our results in
Fig.~\ref{fig:comparison-with-single-well} (a) and (b) for the case of $L=1, 2,
3, 4$.

In Fig.~\ref{fig:comparison-with-single-well} (a), we plot the time
dependence of molecular density for different $L$ with certain atomic interaction strength ($U_a=2g$), and plot the
corresponding dependence of the atom-to-molecule conversion
efficiency on the atomic interaction strength $U_a$ in
Fig.~\ref{fig:comparison-with-single-well} (b).
The
molecular density oscillating periodically with time for any value of
$L$ is displayed in Fig.~\ref{fig:comparison-with-single-well} (a). The larger the number of lattice site, the longer the
oscillation period. From
Fig.~\ref{fig:comparison-with-single-well} (b), we find that the
atom-to-molecule conversion efficiency becomes smaller when
the atomic interaction strength increases.  For the same atomic
interaction strength, the atom-to-molecule conversion
efficiency becomes higher with the increase of the number of lattice
site, which shows that  one can improve the atom-to-molecule
conversion by magnetic lattice even if the atomic interaction
strength is fixed.

In order to confirm the above numerical results, we study the system with analytical method.
For the symmetrical initial state $a_i=1/\sqrt{L}$ and $b=0$, the lattice sites are equivalent, \ie, the values
of $a_i$ at any time does not change with different $i$. Then we can
introduce $A=a_i\sqrt{L}$. Substituting this definition into
Eq.~(\ref{eq:evolution equation of optical lattice}), we can give
the dynamical equation for $A$ and $b$,
\begin{eqnarray}
\label{eq:evolution equation with renormalization}
 i\dot{A}&=&\frac{U_a}{L}|A|^2A+gbA^*,\nonumber\\
 i\dot{b}&=&\frac{g}{2}A^2,
\end{eqnarray}
with the particle conservation law $|A|^2+2|b|^2=1$. Eq.~(\ref{eq:evolution equation with renormalization}) is
similar to the dynamical equation for the atom-to-molecule
conversion system in single well  except for the rescale  of
the atomic interaction strength. From Eq.~(\ref{eq:evolution
equation with renormalization}), we can see that equally distributing  the same atomic BEC into
magnetic lattice can reduce the effective interaction strength
between atoms for the case of $J=0$ through comparing with the
case of single well. Additionally, we know that the
existence of the atomic interaction can suppress the
atom-to-molecule conversion, which can be confirmed by
Fig.~\ref{fig:comparison-with-single-well}~(b). So for the same
atomic interaction strength, the existence of deep magnetic lattice can
improve the atom-to-molecule conversion. Because the effective atomic interaction
strength is $U_a/L$£¬the larger the number of
lattice site is, the higher the atom-to-molecule conversion
efficiency will be. We only consider the
symmetrical initial state above. For the asymmetrical initial state, \ie,
the initial  atomic distributions $|a_i(0)|^2$ are not the same, the
existence of deep magnetic lattice can also improve the atom-to-molecule
conversion if the initial phases of $a_i$ are the same.
Time dependence of molecule density for
different values of $L$ and dependence of the
atom-to-molecule conversion efficiency on the atomic interaction
strength with asymmetrical initial state are shown with Fig.~\ref{fig:comparison-with-single-well}~(c) and (d).
The asymmetrical initial states $|a_1(0)|^2=1/4$, $|a_2(0)|^2=3/4$, $|b(0)|^2=0$ for double-well and
 $|a_1(0)|^2=1/4$, $|a_2(0)|^2=1/4$, $|a_3(0)|^2=1/2$, $|b(0)|^2=0$ for triple-well are considered.
However, for the case
of initial $a_i$ with different phases, lattice can not
always improve the atom-to-molecule conversion.
We should note that the phases of the condensates between
lattice sites are well defined in the Mott insulator state for
the existence of atom-to-molecule conversion.
The phases of the condensates between lattice sites
remain unchanged if they are the same initially.
However, the phases of the atomic condensates between lattice sites will evolve over time
if they are different initially.

\section{Summary and discussion }\label{sec:summary}
In this paper, we have studied the atom-to-molecule conversion in a magnetic lattice.
For shallow lattice, where the atomic tunnelling strength $J$
is unneglected, we studied the
effect of tunnelling and interaction strengths of the
system on the atom-to-molecule conversion by taking double-well as an example.
We gave the dependence of the atom-to-molecule conversion efficiency on the tunnelling strength of atoms and the atomic interaction for the cases of double-well and triple-well.
The general critical points for both cases were found.
We also showed that atoms could be converted into molecules completely if one chose appropriate parameters, \ie, the parameters on
the red solid lines in Fig.~\ref{fig:critical_behavior} (a) and (b) while not completely with other parameters.
The analyses of fixed points and the energy contour were given to confirm our results.
For deep
lattice, where the atomic tunnelling strength $J$ is neglected, we showed that if the initial phases of BECs in different lattice sites
were equal, the existence of lattice site improved the
atom-to-molecule conversion. Considering a symmetrical initial state, we showed that the larger the number of lattice site is,
the higher the atom-to-molecule conversion efficiency could be reached. We
also confirmed our results by an analytical method with the help of
the redefinition of $a_i$. In a word, we gave some suggestions on how to obtain higher conversion efficiency with suitable relation between tunnelling and interaction strengths.

The work is supported by NSFC Grant No. l10674117, No. 11074216 and
partially by PCSIRT Grant No. IRT0754.

\end{document}